\documentclass[aps,prd,superscriptaddress,floatfix]{revtex4}
\usepackage{amsmath,amssymb,url}
\usepackage{rotating}
\usepackage{feynmf}
\usepackage{hyperref}
\usepackage[usenames]{color}

\def\ie{{\it i.e.\;}}
\def\eg{{\it e.g.\;}}

\newcommand{\ope}{\mathcal{O}}
\newcommand{\mv}{m_V^2}

\def\mkk{m_\text{KK}}
\def\mstar{M_*}
\def\mpl{M_\text{Pl}}

\def\mtrans{\Lambda_T}

\begin{document}

\date{\today}
\title{Asymptotically Safe Gravitons in Electroweak Precision Physics}

\author{Erik Gerwick}
\email{E.gerwick@sms.ed.ac.uk}
\affiliation{SUPA, School of Physics and Astronomy, 
             The University of Edinburgh, Scotland, UK}
\begin{abstract}
  Asymptotic safety offers a field theory based UV completion to gravity.  
  For low Planck scales, gravitational effects on low-energy
  precision observables cannot be neglected.  We compute the contribution 
  to the $\rho$ parameter from asymptotically safe gravitons
  and find that in contrast to effective theory, constraints 
  on models with more than three extra dimensions 
  are significantly weakened. The relative size of the trans-Planckian contribution
  increases proportional to the number of extra dimensions.
    
\end{abstract}
\maketitle
%
%
\begin{fmffile}{SM}
%
%
\section{Introduction}
\label{sec:intro}
It is well known that despite the perturbative nonrenormalizability of 
general relativity, it is possible to compute gravitational quantum 
corrections with no assumptions other than those of effective theory~\cite{eft}.  
However, it is also known that power-counting non-renormalizable 
theories may in some cases lead to quantum theories which are finite in 
the UV through non-perturbative dynamics.  One well understood 
case is the Gross-Neveu model in $2+1$ dimensions~\cite{gn}.  There is a 
growing amount of evidence that gravity is another theory featuring non-perturbative 
renormalizability or asymptotic 
safety~\cite{weinberg,reuter_orig,fp_review,more_as,lit_as_saf,fp_extrad,wetterich_eq}.   
The existence of fixed points in higher dimensional gravity \cite{fp_extrad}, 
and their preservation upon coupling to matter \cite{more_as}, suggest that 
extra dimensional models with brane confined matter might also feature
asymptotic safety. 
In this paper we will implement a framework for computing one-loop observables 
in the asymptotic safety scenario and compare  
with previous approaches from effective field theory. This follows a series 
of papers looking at calculable and measurable effects of asymptotically safe
gravity in models with large extra dimensions~\cite{add}, for example at the 
LHC~\cite{fp_pheno,fp_pheno_lp,fp_pheno_hr,prd}.

Recent one-loop computations for the back reaction 
of gravity on the running of the gauge coupling have devoted 
a sizeable effort to 
the issue of gauge dependency in the final result~\cite{pertgrav}.  
Similarly, calculations for precision observables in low scale 
gravity/brane models have produced some debate on gauge 
fixing~\cite{tao_coll,firstew,han_ew,gmuon,noteonew,branons,brane_ew,rad_kk,lpf}.  
It was found that at the amplitude level 
these calculations produce a prohibitively large IR enhancement.  
These effects are gauge dependent and must disappear when 
calculated in proper observables~\cite{branons}.  Since the graviton couples 
universally and ubiquitously, the set of one-loop observables must be 
carefully quantified.  However, our interest is not in the IR spectrum of 
gravity, and thus we will follow Ref.~\cite{branons} and work 
exclusively in the DeDonder gauge where IR divergences are omitted 
from the outset.  The physical interpretation of this gauge is that brane 
fluctuations decouple from the gravitational theory entirely and may 
be studied independently~\cite{sundrum}.  Using the 
DeDonder gauge we will compute corrections to the gauge boson
propagators and find, as expected, that these are sensitive
to the cutoff in both the momentum integral and Kaluza-Klein (KK) summation.  
At this point we will implement asymptotic safety in
an attempt to quantify possible effects occurring at and 
above this scale. 

It was argued previously that in extra dimensional models 
the physical scale identified with the renormalization group 
scale $k$ should be the ($4+n$)-dimensional momentum of 
a KK graviton~\cite{prd}.  For the case of asymptotic safety 
this leads to finite, cut-off independent tree-level scattering 
amplitudes.  We will demonstrate through explicit computation 
that one-loop corrections to gauge boson self energies are also finite in 
this set-up.  Most importantly this means that we have no 
need to apply an explicit UV cutoff. 
 
With a finite calculation for a well defined observable in our 
possession, we would like to explore two questions. 
First, how does the full computation in asymptotic safety compare 
with an effective theory approach.  It is not clear \emph{a priori}  
which will lead to stronger constraints as there are two competing 
effects.  In effective theory the KK integral is 
cut-off separately in the KK mass $\mkk<\Lambda$ and 
the graviton 4-momentum $k<\Lambda$, and
thus (after an analytic continuation) may include modes with
$\sqrt{\mkk^2+k^2}>\Lambda$.  The region outside this circle
is suppressed by UV effects in our framework.  On the other hand, 
modes with $\text{max}(k,\mkk)>\Lambda$ are not included in the 
effective theory computation, while these modes will contribute
in our computation albeit with the UV suppression provided
by asymptotic safety.  The results may then also be compared with 
constraints on extra dimensions from collider~\cite{colliders}, 
astrophysical~\cite{add_astro}, and precision tests of gravity~\cite{gravity_tests}.  
  
Second, we would like to have some sense of the ratio of UV to IR contributions 
as a function of the number of extra dimensions.  We know that tree-level 
amplitudes can become UV dominated for $6$ and more extra  
dimensions~\cite{fp_pheno_lp}. In this paper we will show that at one-loop 
UV domination may begin at 
lower dimensions. 
%
%
%
\section{Self Energy}
\label{sec:amp}
Our task in this section is, starting from the conventional picture of 
large extra dimensions, to compute the self energy amplitudes using 
standard perturbative techniques, and thus reduce the general tensor 
integral into a closed set of scalar integrals.  We will restrict the brane 
dimension to strictly ($3+1$), \ie we  will not use dimensional 
regularization.  Thus, we consider a gravitational theory in ($4+n$)
space-time dimensions with the coupling to matter
\begin{equation}
\frac{32\pi}{\mstar^{2+n}} \, ,
\end{equation}
where $\mstar$ is the fundamental scale of gravity.  The normalization
factor matches the notation of Ref.~\cite{plefka_add}.  For a 
realistic model we require some mechanism which generates the observed 
Planck mass and generically this will involve a volume factor $V_n$ 
of the extra dimensions.  Thus, in the case of a torus compactification
we have the for observed 4-dimensional gravitational coupling
$\mpl^2=(2\pi R)^n \mstar^{2+n}$.  The utility of these conventions
is that the integral and KK sum associated with each graviton loop is
\begin{alignat}{5}
\frac{1}{\mpl^2} \sum_{\text{KK}} \int\frac{d^4k}{(2\pi)^4} =\frac{1}{\mstar^{2+n}}
\int\frac{d^{4+n}q}{(2\pi)^{4+n}} \; ,
\end{alignat}
which is merely a loop integral over the ($4+n$)-dimensional 
momentum.  The sum is shorthand for the individual sums over the KK
occupation numbers. Later, we will need the 4-dimensional brane projection
of $q$, which we denote as $k$.

The gauge fixing and KK reduction is standard~\cite{plefka_add}.  
On the brane we have simply KK gravitons and KK 
scalars in the spectrum, minimally coupled to the Standard Model.  
In the KK reduction, one also in general obtains KK vector 
fields, which do not couple to matter fields at this order in the 
expansion.  
The graviton and scalar propagators are required for our 
computations and are derived by inverting the kinetic terms of the 
Einstein-Hilbert action.  
In the classical regime we expect the scalar part for both to be  
$\Delta_G=(k^2-\mkk^2+i\epsilon)^{-1}$,  but at high energies we anticipate 
different behavior.   

The vertices coupling the scalar and graviton to the brane confined SM can be
obtained by expanding the metric in the massive gauge boson action  
\begin{equation}
g_{\mu\nu}= \sum_{\vec{n}}\eta_{\mu\nu}
\left(1+\frac{16\pi}{\mpl}\phi^{(\vec{n})}\right) 
+ \frac{32\pi}{\mpl} h_{\mu\nu}^{(\vec{n})}+\cdots
\label{eq:metric}
\end{equation}
The sum is over the discrete momenta in the theory.   Implicitly in this
step we have assumed a compact space and enforced 
periodic conditions.  The ellipses stand for terms, in addition to higher 
order in $\mpl^{-1}$, from the vector KK modes and brane fluctuations (branons) 
which do not contribute in our computations~\cite{branons,plefka_add}.

We start with the seagull diagram,
\vspace*{-.2cm}
\begin{alignat}{5} 
\Pi_S(p^2)
&=\;\frac{1}{2}\;\;\; \parbox{25mm}{\begin{fmfgraph*}(70,70)\fmfkeep{boson}
\fmfleft{i} \fmfright{o} \fmf{boson}{i,v} \fmf{double}{v,v}
\fmf{boson}{v,o}
\put(10,60){$h_{\mu\nu}$}
\put(10,25){$p $}
\put(55,25){$p $}
\put(33,50){$q $}
\end{fmfgraph*}} 
\notag \\ 
\vspace{-.6cm}
&=\; 
-\frac{32\pi}{\mstar^{2+n}}\int \frac{d^{4+n}q}{(2\pi)^{4+n}}  
\; \Delta_G\; \frac{3}{2}\left(p^2\eta^{\mu\nu}-p^\mu p^\nu\right) ,
\label{eq:seagull}
\end{alignat}
in agreement with the amplitude for a massless gauge boson.  Here, $p$ 
is the external momentum, $q$ is the bulk loop momentum, and $\Delta_G$ 
the scalar part of the graviton propagator.
Furthermore, the gravi-scalar contribution in four dimensions is proportional to 
the trace of the energy-momentum tensor and thus can only couple to 
the mass term.  However we find in the metric expansion that there is 
no term of order $\mpl^{-2} \phi^2$ and thus the only contribution from 
extra dimensions is Eq.(\ref{eq:seagull}).  We will not simplify 
the loop integral at this point.
The rainbow diagram is 
\vspace*{-.3cm}
\begin{alignat}{5} 
\Pi_R(p^2)
\;\;&=\;\;\
\parbox{20mm}{\begin{fmfgraph*}(70,70)\fmfkeep{boson}
\fmfleft{i} \fmfright{o} 
\fmf{boson}{i,v1} \fmf{boson}{v2,o} 
\fmf{dots,left=.5,tension=0.3}{v1,v2}\fmffreeze
\fmf{boson}{v1,v2}
\put(25,55){$h_{\mu\nu}$}
\put(1,27){$p $}
\put(65,27){$p $}
\put(24,27){$k+p $}
\end{fmfgraph*}}
\quad\;\;\;+\;\;
\parbox{25mm}{\begin{fmfgraph*}(70,70)\fmfkeep{boson}
\fmfleft{i} \fmfright{o} 
\fmf{boson}{i,v1} \fmf{boson}{v2,o} 
\fmf{double,left=.5,tension=0.3}{v1,v2}\fmffreeze
\fmf{boson}{v1,v2}
\put(30,55){$\phi$}
\put(1,27){$p $}
\put(65,27){$p $}
\put(24,27){$k+p $}
\end{fmfgraph*}} 
\notag
\\
&= \quad
\frac{32\pi}{\mstar^{2+n}} 
\int \frac{d^{4+n}q}{(2\pi)^{4+n}}  
\; \Delta_G \; \frac{A^{\mu\nu}(k,p)}
{(k+p)^2-\mv} , 
\label{eq:rainbow}
\end{alignat}
where $A_{\mu\nu}$ contains additionally terms proportional to $\mv$ and
$m_V^4$.  The exact form is given in the Appendix \ref{append}.  We 
have also introduced $k$, the brane only component of the full loop momentum.   
At this point we could introduce Feynman parameters and perform a 
cut-off or dimensional regularization.  We expect that the leading UV divergences 
would cancel with similar terms from the seagull diagram.  However, the 
amplitude would retain power and logarithmic sensitivity to the cut-off scale 
\cite{branons}.  In both cases the amplitude is UV sensitive, and the effects of 
quantum gravity are explicitly cut-off.\medskip

As noted in the introduction, the fixed point scaling which 
we will employ leads to a finite amplitude.  Therefore,  we will 
tensor reduce this amplitude without making specific reference to the 
regulator.  Additionally, the amplitude in Eq.(\ref{eq:rainbow}) can be reduced to 
scalar integrals without any knowledge of the higher dimensional theory.
Using Lorentz invariance on the brane, we project the tensor integral 
onto the scalar integrals 
\begin{alignat}{5}
\tilde{A}_0&= 
\int \frac{d^{4+n}q}{(2\pi)^{4+n}} 
\; \Delta_G 
\notag \\
\tilde{B}_0&= 
\int \frac{d^{4+n }q}{(2\pi)^{4+n }} 
\; \Delta_G \; \frac{1}{(k+p)^2-\mv}
\notag \\
p^2 \tilde{B}_1&= 
\int \frac{d^{4+n }q}{(2\pi)^{4+n }} 
\; \Delta_G \; \frac{p\cdot k}{(k+p)^2-\mv}
\notag \\
p^2 \tilde{B}_2&= 
\int \frac{d^{4+n }q}{(2\pi)^{4+n }} 
\; \Delta_G \; \frac{k^2}{(k+p)^2-\mv}
\label{eq:scalarint}
\end{alignat}
The notation is meant to evoke the similarity to the normal
basis integrals obtained in Passarino-Veltman reduction~\cite{veltman}.  
The seagull contribution simply reads
\begin{equation}
\Pi_S(p^2)= 
-\frac{32\pi}{\mstar^{2+n }} 
\frac{3}{2} \; 
\left( p^2\eta^{\mu\nu}-p^\mu p^\nu \right) 
\tilde{A}_0 \; ,
\end{equation}
while the rainbow self energy in terms of the basis integrals is
\begin{alignat}{4}
&\Pi_R(p^2)= 
\frac{16\pi}{\mstar^{2+n }} \times
\left( (p^2\eta^{\mu\nu}-p^\mu p^\nu) \; 
\left[
12 \mv\left(\tilde{B}_1 + \tilde{B}_0 \right)
  \right.\right.
\notag 
\\
&\left.\left.  +3p^2 \tilde{B}_2 + 16p^2 \tilde{B}_1 + 8p^2 \tilde{B}_0 \right]
+\eta^{\mu\nu} m_V^4 
 \left[ 10+\frac{n }{2+n }\right] \tilde{B}_0
 \right).
 \label{raintot}
\end{alignat}
It should be noted that we have not taken into account brane fluctuations,
coupling proportional to the brane tension $\tau$.  It is not clear how one 
would incorporate this parameter within the framework of asymptotic safety.  
Moreover, we are physically justified to neglect this contribution if we restrict 
ourselves to rigid branes or more precisely the energy hierarchy described  
in Ref.~\cite{plefka_add}.  
%
%
\section{Momentum Integration}
\label{sec:mom}

Quantum corrections to a massless propagator can be simply accounted 
for by the wave function renormalization $Z(\mu)$.  In particular, we define 
the renormalization group improved scalar propagator
\begin{equation}
\Delta(q)=\frac{1}{Z} \; \frac{1}{q^2} \; ,
\label{eq:renorm}
\end{equation}
for a massless graviton.  The specific form of $Z^{-1}$ encodes, in 
general in a non-perturbative way, the quantum effects in the 
theory.  For practical purposes we may only calculate the 
fixed point behavior of $Z$ in certain approximations, \eg with a 
truncated action or at fixed order in perturbation theory.  For this 
study we will use three approximations for $Z^{-1}$ motivated by 
asymptotic safety.
First, we define the linear~\cite{prd}
\begin{equation}
\label{linear}
\Delta^\text{linear} = 
\dfrac{1}{q^2} \left[ 
1+\dfrac{\mu^{n+2}}{\mtrans^{n+2}}
\right]^{-1} \; ,
\end{equation}
and quadratic approximation
\begin{equation}
\label{quadnew}
\Delta^\text{quadratic} = 
\frac{1}{q^2} \left[ 
\sqrt{1+\left(\frac{\mu^{n+2 }}{2\mtrans^{n+2 }}\right)^2} 
-\frac{\mu^{n+2 }}{2\mtrans^{n+2 }} 
\right] \; .
\end{equation}
The numerical value of $\mtrans$ is numerically related to $g^*$, the fixed
point value of the dimensionless coupling, which itself carries gauge and 
cut-off dependency.  Therefore, for our purposes $\mtrans$ is treated as 
input although related to the fundamental scale $\mstar$ by a parameter 
of $\mathcal{O}(1)$.  The factor two in Eq.(\ref{quadnew}) is convenient 
when matching with the one-loop perturbative result \cite{eft}, and thus is left explicit 
rather than being absorbed into the scale $\mtrans$.  

\medskip 

As opposed to the quenched approximation~\cite{lit_as_saf}
\begin{equation}
   \Delta^\text{quenched} = \left\{
     \begin{array}{lr}
       \dfrac{1}{q^2} \qquad \qquad &  q^2 < \mtrans^2\\[4mm]
       \dfrac{1}{q^2}  \; \dfrac{\mtrans^{2+n }}{|q|^{2+n }}
        &  q^2 > \mtrans^2
       \end{array}
   \right. \; ,
   \label{eq:gprop}
\end{equation} 
the graviton propagator in (\ref{linear}) and (\ref{quadnew}) does 
not contain additional poles and thus has only the simple poles 
corresponding to the classical propagator.  These modified propagators 
falls off with sufficient power suppression for large values of $q_0$,  
so the contour integral can be Wick rotated as usual.  

Finally, we note that recent studies indicate that the running of the gauge 
couplings induced by the gravitational coupling 
vanish once a consistent regulator is chosen which respects the appropriate 
symmetries \cite{lpf}.  Therefore, while the graviton propagator is strongly altered above $\mtrans$, we 
assume that the gauge boson propagator remains classical.

\subsection{Finiteness}

We can heuristically show the finiteness
of our calculation in the asymptotic safety framework.  We apply
both the quadratic (\ref{quadnew}) and quenched (\ref{eq:gprop}) approximations to account 
for the large anomalous dimension in the UV.
First of all, we consider the UV portion of the seagull integral in Eq.(\ref{eq:seagull}), 
assuming the wave-function renormalization in the the quenched approximation 
Eq.(\ref{eq:gprop}).  The momentum integral is then trivial
\begin{alignat}{5}
\Pi_{S}^{\text{quenched}} \sim\int_{\mtrans}^{\Lambda_\text{UV}}\frac{dq}{q} 
= \log \frac{\Lambda_\text{UV}}{\mtrans} \; . 
\label{eq:seaquench}
\end{alignat}
For the rainbow diagram we first write the brane momenta as the 4-dimensional 
projection of the ($4 + n $)-dimensional momentum \ie $k^2=q^2\chi^2$.  We will only display
an explicit form of $\chi$ at a later point, since it is not required for showing
UV finiteness.   Considering small values of the external momenta, 
the leading divergence is
\begin{alignat}{5}
\Pi_R^{\text{quenched}}  \sim\int_{\mtrans}^{\Lambda_\text{UV}}\frac{dq}{q}  \;
\frac{q^2 \chi^2}{q^2\chi^2+\mv}= 
\log \frac{\Lambda_\text{UV}}{\mtrans+\cdots }\; .
\label{eq:rainquench}
\end{alignat}
Once we properly perform the tensor reduction the seagull and rainbow 
diagrams acquire pre-factors $(3 /2) (p^2\eta^{\mu\nu}-p^\mu p^\nu)$
with opposite signs, and the sum  of Eq.(\ref{eq:seaquench}) and Eq.(\ref{eq:rainquench}) 
becomes independent of $\Lambda_\text{UV}$ and hence finite.\medskip

It is instructive as well to consider the same integrals for the quadratic approximation.  
For the seagull we find
\begin{alignat}{5}
\Pi_S^{\text{quadratic}} \sim
\frac{\mtrans^{n+2}}{(n+2)}  \sinh^{-1}
\frac{ \Lambda_\text{UV}^{n+2}}{2 \mtrans^{n+2}} \; .
\label{eq:seaquad}
\end{alignat}
We obtain the term in $\Lambda_\text{UV}$  by expanding Eq.(\ref{eq:seaquad}) and 
find the same leading logarithmic dependence on 
 $\Lambda_\text{UV}$ as in the quenched case.  The rainbow we evaluate using 
 the techniques described above, 
and when summed with Eq.(\ref{eq:seaquad}) the result is again finite.  
Had either of these computations provided sub-leading divergences of any type, 
we would not have a sufficiently regulated theory, so the fact that the only divergences 
are logarithmic serves as evidence of finiteness at one-loop.\medskip  

We have not mentioned terms in the amplitude proportional to the $\mv$ and $m_V^4$.
By power counting the later cannot produce a divergence, while the former can at 
worst admit a term proportional to $p\cdot k$ and not $k^2$.  This can be seen by 
examining the momentum structure of the vertices.  These  
terms are individually finite in asymptotically safe gravity.\medskip

The lack of sensitivity to $\Lambda_\text{UV}$ in the self energy
amplitude is of course only valid for the DeDonder gauge and will not
be true for any other choice.  However, also in a less appropriate
gauge any physical observable must be independent of
$\Lambda_\text{UV}$.  To emphasize this point we outline the 
related computation in unitary gauge.  The amplitude for the seagull 
diagram is
\begin{alignat}{5}
\Pi_S^\text{unitary}=
\frac{32\pi}{\mstar^{2+n}}\int &\frac{d^{4+n }q}{(2\pi)^{4+n }}  
\; \Delta_G 
\; \left(p^2\eta^{\mu\nu}-p^\mu p^\nu\right) \times
\notag
\\
&\left( \frac{3}{2}
      -\frac{3}{4}\frac{k^2}{k_T^2}
      +\frac{1}{16}\frac{k^4}{k_T^4}
\right) \; ,
\end{alignat}
with a similar, albeit lengthy, expression for the rainbow.
There are no higher UV divergences since upon substitution 
as described in the following sections the additional terms 
are proportional only to the ratio of the angular parts, with no 
$q$ dependence.  However, the cancellation of the additional 
logarithmic terms must also be checked.  For genuine physical 
observables, such as the one presented later in Section~\ref{sec:res}, 
this cancellation does indeed take place and the result is finite.  
\subsection{Warm-up in $2+1$ dimensions}

To illustrate the geometrical picture of our loop integral we consider the 
same integral with a (Euclidean) 2 dimensional brane and a single extra dimension.
In momentum space we can easily picture the integral over brane and bulk momenta as a three
dimensional integral in terms of spherical polar coordinates.  For a rainbow like diagram
we have
\begin{alignat}{5}
\Pi_{2+1}(p^2) 
&= \int d^3 q \; \frac{1}{q^2 }\frac{k_2^2}
{(k_2 +p)^2+\mv} \notag \\
&= \int d q  \; q^2 \int_{0}^{\pi}d\theta \sin\theta 
\int_{0}^{2\pi} d\phi 
\notag \\
&\frac{1}{q^2 }
\frac{q^2\sin^2\theta}{q^2\sin^2\theta+2|p||q|\sin\theta\cos\phi+ 
p^2+\mv} \; , 
\label{eq:twoplusone}
\end{alignat}
where we denote the 2-dimensional brane analogy of the full
momentum by $k_2$. The spherical coordinates we 
orient such that $p$ is located at $\phi=0$ and
the dot product occurs between the brane projection $|q|\sin\theta$ with the 
brane external momentum $p$.  This integral is  
evaluated numerically, or in this simple case analytically over the angular
coordinates.  The full result in ($4+n $) dimensions is  
merely a higher dimensional generalization of Eq.(\ref{eq:twoplusone}).
  
\subsection{Result in $4+n $ dimensions}

Having established the finiteness of our computation in both the quenched 
and quadratic approximation, we can numerically evaluate the rainbow and seagull diagrams.
It is clear at this point that we would like to perform the ($4+n $)-dimensional momentum 
integral over a radial coordinate $|q|$, but the loop momentum in the gauge boson 
propagator depends on only the brane projection of $|q|$.  Thus our integral in 
($4+n $) dimensions must retain some angular dependence as the 
bulk and brane momentum are not interchangeable in the rainbow diagram.  

A straightforward solution is to define a ($4+n $) dimensional Euclidean vector 
$q=(k_T, k_4)$ in polar coordinates.  Treating the brane 
momentum as the last entries in this ($4+n $)-vector allows us to make the variable 
change $k_4^2 + k_T^2 \equiv q^2$ which in turn requires the projection 
\begin{equation}
k_4^2 \equiv q^2 \, \sin^2 \phi_1 \, \sin^2 \phi_2 \, \cdots \, \sin^2 \phi_n 
      \equiv q^2 \chi^2 \; ,
\label{eq:varchange}
\end{equation}
where the last step is simply a shorthand notation.
The physical interpretation of the angular variables should be clear as they 
measure the $4$-dimensional projection of $q$.   The configuration with all
$\phi_i=\pi/2$ corresponds to momenta confined to the brane, and the corresponding integral 
is the conventional 4-dimensional case.  This simplifies our loop integrals and is 
easily evaluated numerically.  Under the coordinate change the measure in our 
($4+n $)-dimensional loop integration should be replaced as
\begin{alignat}{5}
\int\frac{ d^{4+n }  q}{(2\pi)^{4+n }}
= 
\frac{\pi}{(2\pi)^{4+n } }&\int dq \,q^{3+n } 
\int_0^{\pi} d\phi_1 \sin^{2+n }\phi_1 \cdots
\notag \\
&
\int_0^{\pi} d\phi_n  \sin^3 \phi_n 
\int_0^{2\pi} d\phi_{3+n }.
\label{eq:measure}
\end{alignat}
The pre-factor $\pi$ in front is the result of integrating over the 2-additional 
brane angular coordinates $\phi_{1+n}$ and $\phi_{2+n}$ which our amplitude 
does not depend on.  For the angular measure related to $\chi$ we write 
$\int d\chi$, so the basis integrals from Eq.(\ref{eq:scalarint}) with the 
renormalization group improved scalar propagator in Eq.(\ref{eq:renorm}) 
become
\begin{alignat}{5}
\tilde{A}_0 =\frac{2}{(2\pi)^{4+n }}\frac{\pi^{n/2+2}}{\Gamma[n/2+2]}
\int dq\, q^{n+1} Z^{-1}
\end{alignat}
\begin{alignat}{5}
\tilde{B}_0 &=
\frac{\pi}{(2\pi)^{4+n }} \int dq \,q^{3+n }\int  d\chi \int_0^{2\pi} d\phi_{3+n}
\notag \\ & \frac{Z^{-1}}{q^2}
\; \frac{1}{q^2\chi^{2}+2|p||q|\cos\phi_{3+n }\chi+(p^2+\mv)}
\end{alignat}
\begin{alignat}{5}
p^2 \tilde{B}_1 &=
\frac{\pi}{(2\pi)^{4+n }} \int dq \,q^{3+n }\int  d\chi \int_0^{2\pi} d\phi_{3+n}
\notag \\ & \frac{Z^{-1}}{q^2}
\; \frac{|p||q|\cos\phi_{3+n }}
 {q^2\chi+2|p||q|\cos\phi_{3+n }+(p^2+\mv)\chi^{-1}} 
\end{alignat}
\vspace{-.5cm}
\begin{alignat}{5}
\notag \\
p^2 \tilde{B}_2 &=
\frac{\pi}{(2\pi)^{4+n }} \int dq \,q^{3+n }\int  d\chi \int_0^{2\pi} d\phi_{3+n}
\notag \\ & \frac{Z^{-1}}{q^2+2|p||q|\cos\phi_{3+n }\chi^{-1}+(p^2+\mv)\chi^{-2}}
\end{alignat}
Using these basis integrals we can numerically compute the
gauge bosons self energies in Eq.(\ref{eq:seagull}) and Eq.(\ref{raintot}). 
The sum of the two integrals does not depend on 
$\Lambda_\text{UV}$.
In the next section we will see how this method of computing the KK integral 
compares with an effective theory analysis.
%
%
%
\section{Results}
\label{sec:res}

\begin{figure}[t]
\includegraphics[width=0.42 \textwidth]{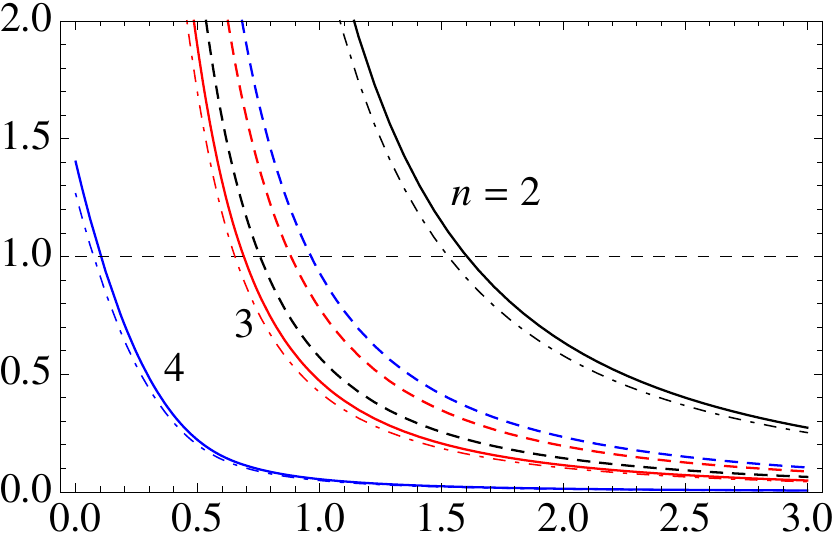}
\put(-174,100){$\bar{\epsilon}\times 10^{3}$}
\put(0,0){$\mstar$}
\linethickness{.3mm}
\put(-61,107){\color{black}\line(2,0){2.5}}
\put(-57,107){\color{black}\line(2,0){2.5}}
\put(-53,107){\color{black}\line(2,0){2.5}}
\put(-45,105){EFT}
\put(-61,97){\color{black}\line(2,0){10}}
\put(-45,95){FP}
\hspace{0.1 \textwidth}
\includegraphics[width=0.42 \textwidth]{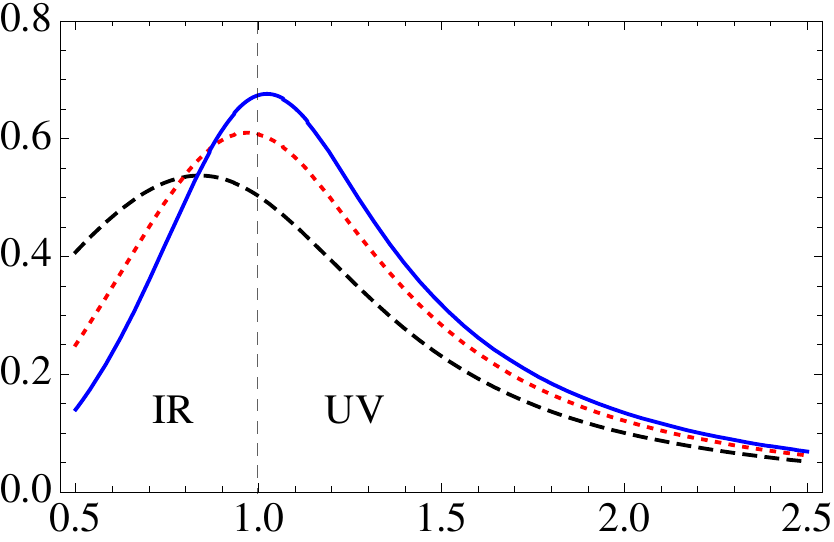}
\put(-170,100){$\dfrac{1}{\bar{\epsilon}}\dfrac{d\bar{\epsilon}}{d\Lambda}$}
\put(0,0){$\dfrac{\Lambda}{\mstar}$}
\linethickness{.3mm}
\put(-61,107){\color{black}\line(2,0){2.5}}
\put(-57,107){\color{black}\line(2,0){2.5}}
\put(-53,107){\color{black}\line(2,0){2.5}}
\put(-45,105){$n=2$}
\put(-61,97){\color{red}\line(2,0){1}}
\put(-59,97){\color{red}\line(2,0){1}}
\put(-57,97){\color{red}\line(2,0){1}}
\put(-55,97){\color{red}\line(2,0){1}}
\put(-53,97){\color{red}\line(2,0){1}}
\put(-45,95){$n=3$}
\put(-61,87){\color{blue}\line(2,0){10}}
\put(-45,85){$n=4$}
\caption{Left: contribution to $\bar{\epsilon} \simeq 1- \rho$ from
  extra dimensions computed in asymptotically safe gravity in linear (\ref{linear}) (dot-dashed), 
  quadratic (\ref{quadnew}) (solid) and in effective theory (dashed)~\cite{branons}. Values above 
  the dashed horizontal curve are in tension with data.  For all asymptotic safety curves we have 
  taken $\mstar = \mtrans $ and in the EFT case we have $\mstar = \Lambda_{\text{eft}}$. 
  Right: contribution to
  $\bar{\epsilon}$ as a function of the highest momentum mode $q < k_0$ in
  the ($4+n$)-dimensional integration.  We see the approximate pattern of Eq. (\ref{eq:simpint}) in
  the ratio of IR to UV contributions.  }
\label{fig:res}
\end{figure}
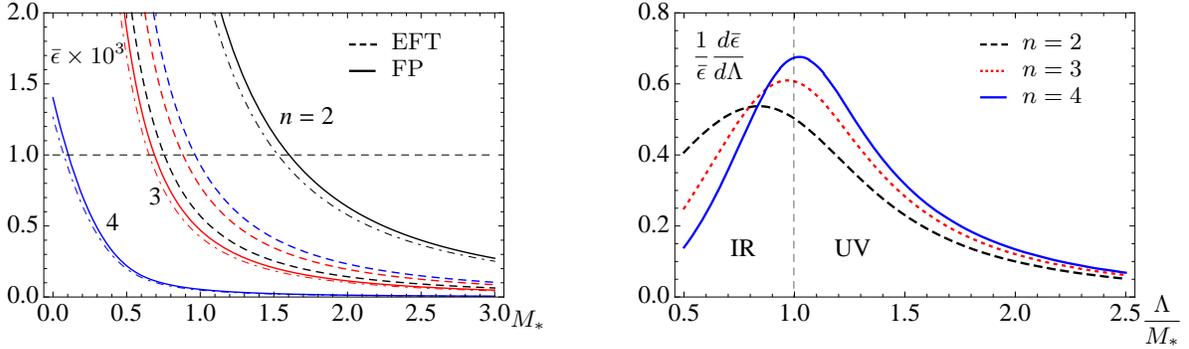

In order to constrain new physics based on precision measurements it
is typical to use either the oblique parameter set
$\{S,T,U\}$~\cite{stu} or the $\epsilon$ parameterization~\cite{epsp}.
However, the former is not a good parameter set for gravitons, which
in general will modify more than just gauge boson self-energies.  The
later can be related to the weak mixing angle $s_0, c_0$ to define the 
leading corrections
\begin{equation}
\rho - 1 \simeq \bar{\epsilon} = \epsilon_1-\epsilon_2-\frac{s_0^2}{c_0^2}\epsilon_3 \; ,
\label{rhodef}
\end{equation}
quantifying the violation of custodial symmetry~\cite{wells_tasi,book_ew}. The individual observables $\epsilon_1, \epsilon_2$ 
and $\epsilon_3$ in the context of gravity require additional computations beyond 
the self-energies, but the specific linear combination 
\begin{equation}
\bar{\epsilon}=\frac{\Pi(m_W^2)}{m_W^2}-\frac{\Pi(m_Z^2)}{m_Z^2}
\end{equation}   
has to be smaller than roughly $10^{-3}$, assuming a light Standard
Model Higgs boson~\cite{working_group}. The exact central value
depends on the Higgs mass, but the uncertainty of $\bar \epsilon$ also
ranges around $10^{-3}$, which makes this value a generic upper bound
on new physics effects.  The effective theory result for gravity
contributions~\cite{branons} in our notation reads
\begin{equation}
\Delta \rho \simeq 
\Delta \bar{\epsilon} = \frac{s^2  m_Z^2}{\mstar^2}
\left( \frac{\Lambda_{\text{eft}}}{\mstar} \right)^n 
\frac{1}{\Gamma(2+n /2)} \; 
\frac{5(8+5n )}{48 \pi^{2-n /2}} \; ,
\label{eq:eft}
\end{equation}
for $m_\text{KK} \gg m_Z$.  Here, $\Lambda_{\text{eft}}$ is the effective 
theory cut-off scale.  In Figure~\ref{fig:res}
we compare this result to our computation as described in the previous
section.  The numerical results are similar for $n =3$, where
fundamental Planck masses below $M_* \lesssim 1$~TeV are forbidden.   In
asymptotically safe gravity the limits are significantly weakened when we
increase the number of extra dimensions, which is not the case for the
effective theory computation in Eq. (\ref{eq:eft}). In other words, once we take 
the higher dimensional quantum effects seriously there are essentially no limits
on large extra dimensions with $n>3$ from electroweak precision data.

The explanation for this discrepancy is the following: in effective theory, the
amplitudes are dominated by modes with both $k$ and $\mkk$ near the
cut-off, \ie in the corner of a square in the $k$ vs $\mkk$ plane. However, in our picture of asymptotic safety with 
$|\mu|=\sqrt{k^2+\mkk^2}$ this region is outside the central circle which means it is
suppressed.  For higher numbers of extra dimensions this
effect is more pronounced, as shown by the lessening contribution to
$\bar{\epsilon}$ as $n$ increases.

To confirm our observation of an increasing impact from the UV regime we
need to study the distribution of the momentum modes contributing to
$\bar{\epsilon}$.  More specifically, we would like to know the size of the 
contribution to the observable for UV momentum modes with $|q|>\mtrans$.
In the quenched approximation we can estimate this fraction  from
Eq.(\ref{eq:gprop}).  Neglecting terms of order $m_V/\mtrans$ the leading 
momentum integrals are of the form
\begin{alignat}{5}
\bar{\epsilon} \Bigg|_\text{IR}
&\sim \int_0^{\mtrans} dq \; q^{3+n } \; \frac{1}{q^2}\; 
\frac{1}{q^2+\cdots}
\approx  \frac{\mtrans^{n }}{n }
\notag \\
\bar{\epsilon} \Bigg|_\text{UV}
&\sim \int_{\mtrans}^{\infty} dq \; q^{3+n } \; \frac{\mtrans^{2+n }}{q^{4+n }} \; 
\frac{1}{q^2+\cdots}
\approx 
\frac{\mtrans^{n }}{2} \; .
\label{eq:simpint}
\end{alignat}
This expansion in 
$m_V/\mtrans$ is well validated, for example compared with LHC tree 
level virtual graviton exchange where $\sqrt{\bar{s}}/\mtrans$ 
can easily become $\ope(1)$~\cite{fp_pheno}.  
For $n  = 2$ approximately $50\%$ of the combined integral comes from
the quantum gravity regime, and for a larger number of extra
dimensions this fraction increases.  Numerically, 
Figure~\ref{fig:res2} shows that the exact results follow the pattern
delineated in Eq.(\ref{eq:simpint}). The UV region becomes the
dominant contribution already for $n >2$ extra dimensions.

This can be contrasted with the case of LHC tree level graviton
exchange, where in an identical limit the UV graviton contribution
becomes dominant for $n >6$~\cite{fp_pheno_lp}.  This is not
surprising though as the momentum integral in the tree-level exchange
is over four less directions in momentum space, and thus we see the
rough equivalence of the $n =6$ case in tree-level with the
$n =2$ case at one-loop.

\section{Conclusion}
\label{sec:con}

Asymptotically safe gravity allows us to compute quantum gravitational effects on 
relevant observables. The ultraviolet or quantum gravity regime does not require any
modified treatment or cut-off procedure, \eg to ensure finiteness of our predictions.
In models with large extra dimensions~\cite{add} gravitational effects 
should be measurable, since the fundamental Planck scale does not 
lead to a significant suppression, as compared to other TeV scale physics 
or electroweak loop effects. In this framework, asymptotic safety is
particularly useful because it provides measurable predictions for the 
LHC~\cite{fp_pheno} or, as we have shown in the work, for electroweak 
precision measurements.\medskip 

We first introduced a method for evaluating gravitational loop integrals
in extra-dimensions.  Defining these in terms of the full ($4+n$)-dimensional 
momentum provided a straight-forward regularization scheme.  Similar to the 
usual Passarino-Veltman reduction into scalar one-loop integrals, we defined
and numerically evaluated a set of basis integrals which can be used for a wide class of
observables. In this paper we studied gravitational effects on custodial symmetry in the 
Standard Model, \ie the $\rho$ parameter or $\bar \epsilon$.\medskip

 We find that the bounds from electroweak precision data based on asymptotic safety 
 are roughly equivalent to the direct/indirect
bounds obtained by previous methods for $n=3$ extra dimensions.  For higher numbers
of extra dimensions the bounds from virtual gravitons are irrelevant compared
with more direct measurements. Compared to an effective field theory (or cut-off) prescription
our limits are weaker for more than two extra dimensions. This is consistent 
with the general observation that for  larger numbers of extra dimensions the relative 
contribution from the trans-Planckian (and hence strongly suppressed) regime become 
more and more dominant. No matter what kind of ultraviolet completion of gravity should
be chosen by Nature, this paper shows that it has to be modeled properly and 
quantitatively taken into account.

\subsection*{Acknowledgments}

I am grateful to Tilman Plehn for continuous encouragement, assisting with the 
write up of this work and, along with the Institut f\"ur Theoretische Physik at 
Heidelberg University, for on-going hospitality. Moreover, I would like to 
thank Daniel Litim, Jan Pawlowski and Einan Gardi for helpful discussion 
and theoretical insights.

\appendix
\section{Amplitude}
\label{append}
In Eq.(\ref{eq:rainbow}) we defined $A_{\mu\nu}$ which is given here for convenience.
\begin{equation}
A^{\mu\nu}(k,p)=A_{1}^{\mu\nu}(k,p)+\mv A_{2}^{\mu\nu}(k,p) + m_V^4 
A_{3}^{\mu\nu}
\end{equation}
\begin{alignat}{5}
     A^{\mu\nu}_1(k,p) =
          \frac{1}{2}&\left[- 4p^{\mu}p^{\nu}p^2
          - 4p^{\mu}p^{\nu}p \cdot k
          - p^{\mu}p^{\nu}k^2 \right.
          \notag
          \\
          &
          - 4p^{\mu}k^{\nu}p^2
          - 3p^{\mu}k^{\nu}p \cdot k
          + p^{\nu}k^{\mu}p \cdot k 
          \notag
          \\
          &
           - k^{\mu}k^{\nu}p^2 
          + 8\eta^{\mu\nu}p^2 p \cdot k
          + \eta^{\mu\nu}p^2 k^2
          \notag
          \\
          &
          \left.
          + 4\eta^{\mu\nu}p^4
          + 3\eta^{\mu\nu} (p \cdot k)^2 \right]
\end{alignat}
\begin{alignat}{5}
         A_{2}^{\mu\nu}(k,p) =
           3 (\eta^{\mu\nu} (p\cdot k+p^2) 
	 -  p^{\mu} k^{\nu}-p^{\mu} p^{\nu})
\end{alignat}
\begin{alignat}{5}
         A_{3}^{\mu\nu} =
         \frac{5}{2} \eta^{\mu\nu}
\end{alignat}
%


\end{fmffile}
\end{document}